\documentclass[aps,prb,groupedaddress,showpacs]{revtex4}

\begin{document}


\title{Cooperative effects in Josephson junctions in a cavity in the strong coupling regime}


\author{Marco Frasca}
\email[]{marcofrasca@mclink.it}
\affiliation{Via Erasmo Gattamelata, 3 \\ 00176 Roma (Italy)}


\date{\today}

\begin{abstract}
We analyze the behavior of systems of two and three qubits made by Josephson 
junctions,
treated in the two level approximation,
driven by a radiation mode in a cavity. The regime we consider is
a strong coupling one recently experimentally reached for a single junction.
Rabi oscillations are obtained with the frequency proportional to integer order
Bessel functions in the limit of a large photon number, similarly to the 
case of the single qubit. A selection rule is derived for the appearance of
Rabi oscillations. A quantum amplifier built with a large number of Josephson
junctions in a cavity in the strong coupling regime is also described. 
\end{abstract}

\pacs{74.50.+r, 03.65.Ud, 03.67.Lx, 42.50.Fx}

\maketitle

\section{Introduction}

Quantum computation \cite{qc1,qc2,qc3} promises a large improvement on the ability to execute
demanding algorithms due to the large parallelism involved. Presently, it
is not yet clearly understood how the hardware for a quantum computer should be realized.
Several proposal relying e.g. on ion traps \cite{cz} and NMR \cite{nmr} have been devised but
solid state devices seem to be very promising for this aim. Josephson junctions have been 
largely used both experimentally and theoretically for this goal \cite{jj1,jj2}.

In a couple of recent experiments Nakamura, Tsai and Pashkin \cite{naka1}
and Chiorescu et al. \cite{naka2} 
were able to display the 
behavior of a 
single
Josephson junction strongly coupled to a radiation field and its suitability
for quantum computation. The strong coupling regime has the advantage that can be theoretically described by
a two-level model \cite{fra1,fujii}.

Validity of the two-level model to describe the behavior of Josephson junctions has been proved by
Stroud and Al-Saidi \cite{stro1,stro2}. They considered the case of weak coupling where the so called 
rotating wave approximation does apply \cite{fra2}. They also proved that, in some cases,
a correction term to the model is needed while collapse and revival 
of Rabi oscillations is observed as also
happens to similar models in quantum optics \cite{ebe,schl}. Similarities between solid state 
devices and quantum optics systems are becoming increasingly meaningful \cite{bra}.

The model used by Stroud and Al-Saidi cannot be applied directly to the experiments of the
Nakamura's group. Rather, we need to use the Dicke model \cite{dicke} without any other
approximations than the numbers of radiation modes and two-level systems. The first complete
application of this kind has been given by us \cite{fra1} even if a study of this model
in the strong coupling regime was started by Cohen-Tannoudji and his group \cite{ct}
for a single radiation mode and a two-level atom. In this case all the terms in the
interaction part of the Hamiltonian must be retained.

In our paper \cite{fra1} we were able to obtain the equations for the probability amplitudes
in the strong coupling regime given the proper set of states. Two bands of levels were obtained
and both intraband and interband transitions were shown. In the limit of a large number of
photons in the cavity, the Rabi frequency is proportional to the integer number Bessel
functions. We will recover this result below. Rather interestingly, the square of
the amplitudes of the levels involved in the Rabi oscillations is a Poisson distribution.
Rabi oscillations arise from the crossing of the energy levels of the Dicke model and
appear between macroscopic superposition of charge states
(known in the current literature as Schr\"odinger cat states).

In the Dicke model, two-level systems are coupled by the radiation field. The same can happen
with Josephson junctions \cite{stro3}. So, our aim is to see how collective effects can emerge
when more Josephson junctions are coupled by a cavity field. We will prove that Rabi oscillations
emerge also in this case with the Rabi frequencies proportional to integer order Bessel functions.
A selection rule arises for situations with more qubits as certain transitions are not allowed.
In turn this means that the transition to some qubit configurations cannot be realized.
Then, states involved in the Rabi oscillations are always macroscopic superpositions of
entangled states between the radiation field, that can have a large
number of photons, and the junctions.

Finally, we extend the analysis to the thermodynamic limit showing that, in this case, a
large number of Josephson junctions coupled by a cavity field in the strong coupling 
regime can be used to amplify vacuum fluctuations of the field to a large macroscopic
field, assuming all the junctions in their ground state. Decoherence effects may be
prominent in this case as we will discuss. This effect defines a quantum amplifier (QAMP)
as proposed by us in the literature \cite{fra2,fra3,fra4}.

The paper is structured as follows. In sec.\ref{sec2} we introduce the model describing
a number of Josephson junctions interacting through a cavity field. In sec.\ref{sec3} we
analyze explicitly the cases for one, two and three coupled qubits obtaining the Rabi
frequencies and proving, in the limit of a large number of photons, the proportionality
with integer order Bessel functions. In sec.\ref{sec4} we discuss the thermodynamic limit, that
is we see the physics of a large number of Josephson junctions in a cavity proving that
an amplification of the vacuum fluctuations of the radiation field in the cavity is
obtained. This new device we term QAMP. Finally, in sec.\ref{sec5} conclusions are given.

\section{\label{sec2} Description of the model}

The model we should consider for a
single Josephson junction,
treated in the two-level approximation,
in a cavity field is (here and in the following $\hbar=1$)
\begin{equation}
\label{eq:naka}
    H = \Delta S_z + \omega a^\dagger a +2gS_x(a^\dagger + a)
\end{equation}
being $\Delta$ the separation between the ground and the first excited state in a Cooper
pair \cite{stro1,stro2}, $\omega$ the frequency of the field in the cavity, $g$ the coupling
between the junction and the cavity, $a^\dagger$ and $a$ creation and annihilation operators,
$S_x,S_z={\sigma_x \over 2}, {\sigma_z \over 2}$ con $\sigma_x,\sigma_z$ Pauli matrices.

Eq.(\ref{eq:naka}) is just the Dicke model that we have specialized to a single qubit with
Pauli matrices. This generalizes immediately to any number of qubits putting
\begin{eqnarray}
\label{eq:SxSz}
    S_x &=& {1 \over 2} \sum_{i=1}^N\sigma_{x,i} \\ \nonumber
    S_z &=& {1 \over 2} \sum_{i=1}^N\sigma_{z,i}
\end{eqnarray}
being $N$ the number of qubits.

One may think that eq.(\ref{eq:naka}) differs from the Hamiltonian currently used to
describe Josephson junction qubits in experiments \cite{naka1,naka2} but these Hamiltonians 
are equivalent as we can interchange $\sigma_x$ and $\sigma_z$ with the unitary
transformation $e^{i{\pi\over 4}\sigma_y}$ and what one gets, in the analysis
in the strong coupling regime, is a shift to the energy levels. So, both Hamiltonians give 
rise to the same physics and we can safely work with the Dicke model as currently known.

In order to work out the analysis of the model in the strong coupling regime proper to
the experiments carried out so far, we follow the same approach used in Ref.\cite{fra1}.
We split the Hamiltonian (\ref{eq:naka}) as
\begin{eqnarray}
    H_0 &=& \omega a^\dagger a +2gS_x(a^\dagger + a) \\ \nonumber
    H_1 &=& \Delta S_z
\end{eqnarray}
being $H_0$ the unperturbed Hamiltonian and $H_1$ the perturbation. The unperturbed part 
$H_0$
can be immediately diagonalized by the eigenstates,
as can be directly verified treating $S_x$ as a c-number,
\begin{equation}
    |[n]; S, S_x\rangle = e^{\frac{2g}{\omega}S_x(a-a^\dagger)}|n\rangle|S,S_x\rangle
\end{equation}
being $S_x$ the component of the spin along x axis chosen as the axis of quantization,
$|S,S_x\rangle$ the corresponding eigenstates and $a^\dagger a|n\rangle=n|n\rangle$
and $n=0,1,2,\ldots$ an integer. Energy eigenvalues are given by
\begin{equation}
    E_{n,S_x} = n\omega - \frac{4g^2S_x^2}{\omega}
\end{equation}
and we observe a degeneracy between positive and negative values of the spin components. Then,
we can introduce the unitary evolution operator
\begin{equation}
\label{eq:uf}
    U_F(t) = \sum_{n,S_x}e^{-i\left(n\omega - \frac{4g^2S_x^2}{\omega}\right)t}
    |[n]; S, S_x\rangle\langle [n]; S, S_x|.
\end{equation}
We can apply this unitary operator to the initial Hamiltonian (\ref{eq:naka}) and the
problem is reduced to the Hamiltonian
\begin{equation}
\label{eq:naka1}
    \tilde H = \sum_{m,S_x}\sum_{n,S'_x}
           e^{i\left(m\omega - \frac{4g^2S_x^2}{\omega}\right)t}
           e^{-i\left(n\omega - \frac{4g^2S_x^{'2}}{\omega}\right)t}
           \Delta |[m]; S, S_x\rangle\langle [n]; S, S'_x|
           \langle [m]; S, S_x|S_z| [n]; S, S'_x\rangle.
\end{equation}
The matrix elements can be evaluated by noting that $S_z=\frac{S_+-S_-}{2i}$ and one gets
\begin{eqnarray}
      \langle [m]; S, S_x|S_z| [n]; S, S'_x\rangle &=& \frac{1}{2i}
      \left(
           \delta_{S_x,S'_x+1}\sqrt{S(S+1)-S'_x(S'_x+1)}M^{-}_{mn}\right. \\ \nonumber
          &-&\left.\delta_{S_x,S'_x-1}\sqrt{S(S+1)-S'_x(S'_x-1)}M^{+}_{mn}
      \right)
\end{eqnarray}
where we have put \cite{fra1}
\begin{equation}
\label{eq:Mnm}
      M^{\pm}_{mn}=\langle m|e^{\pm\frac{2g}{\omega}(a-a^\dagger)}|n\rangle=
      \sqrt{\frac{n!}{m!}}e^{-\frac{2g^2}{\omega^2}}
      \left(\pm\frac{2g}{\omega}\right)^{m-n}L_n^{m-n}
      \left(\frac{4g^2}{\omega^2}\right)
\end{equation}
being $L_n^{m-n}$ Laguerre polynomials.

Now the Hamiltonian (\ref{eq:naka1}) can be split in two parts to give a diagonal part
\begin{eqnarray}
\label{eq:h0}
    \tilde H_0 &=& \sum_{n,S_x}\frac{1}{2i}\Delta M_n\left(
           e^{-i\frac{4g^2(2S_x-1)}{\omega}t}
\sqrt{S(S+1)-S_x(S_x-1)}|[n]; S, S_x\rangle\langle [n]; S, S_x-1|\right. \\ \nonumber
           &-&\left.e^{i\frac{4g^2(2S_x+1)}{\omega}t} 
\sqrt{S(S+1)-S_x(S_x+1)}|[n]; S, S_x\rangle\langle [n]; S, S_x+1|\right),
\end{eqnarray}
having set
\begin{equation}
      M_n=M^{\pm}_{nn}=e^{-\frac{2g^2}{\omega^2}}L_n\left(\frac{4g^2}{\omega^2}\right),
\end{equation}
and an off-diagonal part
\begin{eqnarray}
\label{eq:h1}
    \tilde H_1 &=& \sum_{\stackrel{m,n,S_x}{m\neq n}}\frac{1}{2i}\Delta e^{-i(n-m)\omega t}
\left(
           e^{-i\frac{4g^2(2S_x-1)}{\omega}t}M^{-}_{mn}
\sqrt{S(S+1)-S_x(S_x-1)}|[m]; S, S_x\rangle\langle [n]; S, S_x-1|\right. \\ \nonumber
           &-&\left.e^{i\frac{4g^2(2S_x+1)}{\omega}t} M^{+}_{mn}
\sqrt{S(S+1)-S_x(S_x+1)}|[m]; S, S_x\rangle\langle [n]; S, S_x+1|\right).
\end{eqnarray}

In order to obtain the equations for the probability amplitudes, we diagonalize
Hamiltonian (\ref{eq:h0}) obtaining the energy eigenvalues, the eigenstates and
the geometrical part of the phase as we can have time-dependent eigenstates,
in agreement to the general approach outlined in Ref.\cite{fra1,fra5}. Then, by
interaction picture, the amplitude equations are obtained using Hamiltonian (\ref{eq:h1}).

In the analysis we pursue in the following we will see that Hamiltonian (\ref{eq:h0}) has
eigenvalues $s_x\Delta M_n$ having $s_x$ the same values of the spin projection $S_x$ and
one can have transitions between the eigenstates
only if the selection rule $\Delta s_x = 0, \pm 1$ holds. We recognize that each level
$s_x$ develops a band of levels numbered by integer $n$, so the selection rule provides
intraband $(s_x=\tilde s_x)$ and interband $(s_x\neq\tilde s_x)$ permitted transitions 
at resonance respectively.

In the limit of a large number of photons we can show, in all cases below, that the
Rabi frequencies are proportional to integer order Bessel functions. This can be
accomplished using the relation\cite{fra1}
\begin{equation}
\label{eq:Ja}
      J_\alpha(2\sqrt{nx})=e^{-\frac{x}{2}}
      \left(\frac{x}{n}\right)^{\frac{\alpha}{2}}L_n^\alpha(x)
\end{equation}
that holds in the limit of $n$ going to infinity.

\section{\label{sec3} Analysis of the model}

We apply the procedure outlined in the previous section to one, two and three qubits
to see in details the physics of the model. The one qubit case has been already
discussed in literature\cite{fra1} but we present it here again to show the way our method works
to make this paper self-contained.

\subsection{One qubit}

In the one qubit case we have $S_x={1\over 2},-{1\over 2}$ being $S={1\over 2}$. The diagonal
part of the Hamiltonian wil be
\begin{equation}
\label{eq:h012}
    \tilde H_0 = \sum_n\frac{1}{2i}\Delta M_n\left(
    \left|[n];{1\over 2},{1\over 2}\right\rangle\left\langle [n]; {1\over 2},-{1\over 2}\right|
    - \left|[n]; {1\over 2}, -{1\over 2}\right\rangle\left\langle [n]; {1\over 2},{1\over 2}\right|\right),
\end{equation}
that is time-independent in this case. Similarly, we have the off-diagonal part
\begin{equation}
\label{eq:h112}
    \tilde H_1 = \sum_{\stackrel{m,n}{m\neq n}}\frac{1}{2i}\Delta e^{-i(n-m)\omega t}
     \left(M^{-}_{mn}
    \left|[m];{1\over 2},{1\over 2}\right\rangle\left\langle [n]; {1\over 2},-{1\over 2}\right|
    - M^{+}_{mn}\left|[m]; {1\over 2}, -{1\over 2}\right\rangle
    \left\langle [n]; {1\over 2},{1\over 2}\right|\right).
\end{equation}
This operator is Hermitian as can be verified using the fact that sum indexes are dummy.

Hamiltonian (\ref{eq:h012}) can be immediately diagonalized with eigenvalues 
$E_{n,s_x}=s_x\Delta M_n$ being $s_x={1\over 2},-{1\over 2}$ corresponding to the eigenstates
\begin{equation}
\label{eq:h0st}
     |n; s_x\rangle = \frac{1}{\sqrt{2}}\left(
\left|[n];{1\over 2},-{1\over 2}\right\rangle
-\frac{i}{2s_x}\left|[n];{1\over 2},{1\over 2}\right\rangle
\right)
\end{equation}
that are not eigenstates of the spin projection $S_x$. Then, we can write the equation for the
probability amplitudes by looking for a time-dependent solution in the form
\begin{equation}
\label{eq:ans}
     |\psi(t)\rangle = \sum_{k,s_x}e^{-iE_{k,s_x}t}c_{k,s_x}(t)|k; s_x\rangle 
\end{equation}
where no contribution enters due to geometrical phases as Hamiltonian (\ref{eq:h012}) is
time-independent. The Schr\"odinger equation takes the form
\begin{equation}
\label{eq:eqS}
     i\dot c_{m,\tilde s_x}(t) = \frac{\Delta}{2}\sum_{n,s_x}
     e^{-i[E_{n,s_x}-E_{m,\tilde s_x}+(n-m)\omega]t}
     (\tilde s_x M^{-}_{mn} + s_x M^{+}_{mn}) c_{n,s_x}(t).
\end{equation}
We recognize the resonance conditions
\begin{equation}
     E_{n,s_x}-E_{m,\tilde s_x}+(n-m)\omega = 0
\end{equation}
with $\Delta s_x=0$ for intraband transitions and $\Delta s_x=\pm 1$ for interband transitions as said above.
These resonance conditions correspond to crossing of the energy levels of the Dicke model. Energy levels are given 
by \cite{ene}
\begin{equation}
     E_{n,s_x} = n\omega - \frac{4s_x^2g^2}{\omega} + s_x\Delta M_n
\end{equation}
where we can recognize the band structure and the degeneracy for $s_x=\pm{1\over 2}$
that is removed by the last term.
Then, resonance conditions can be straightforwardly interpreted
as crossings of energy levels. Such crossings produce Rabi oscillations as experimentally observed.

Finally, let us check the Rabi frequencies for the single qubit case. It is not difficult to realize that we have
for intraband transitions 
$\Delta s_x=0$ from eq.(\ref{eq:eqS}), assuming the rotating wave approximation due to the resonant terms,
\begin{equation}
     {\cal R}_1 =
     \frac{1}{2}\Delta\left(M^{+}_{mn} + M^{-}_{mn}\right)=
      \Delta \left\langle m\left|\cosh\left[\frac{2g}{\omega}(a-a^\dagger)\right]\right|n\right\rangle
\end{equation}
and for interband transitions 
$\Delta s_x=\pm 1$
\begin{equation}
     {\cal R'}_1 =
     \frac{1}{2}\Delta\left(M^{+}_{mn} - M^{-}_{mn}\right)= 
     \Delta \left\langle m\left|\sinh\left[\frac{2g}{\omega}(a-a^\dagger)\right]\right|n\right\rangle.
\end{equation}
Now, we can use eq.(\ref{eq:Mnm}) to show that transitions with states differing by an odd number of photons,
$m=n+2N+1$ are permitted for intraband transitions and even number of photons are involved $m=n+2N$ in interband transitions \cite{fra1}. Using Sterling formula and eq.(\ref{eq:Ja}) one gets, in the limit of Fock states with a large number of photons,
\begin{equation}
    {\cal R}_1\approx\Delta J_{2N+1}\left(\frac{4\sqrt{n}g}{\omega}\right)
\end{equation}
and
\begin{equation}
    {\cal R'}_1\approx\Delta J_{2N}\left(\frac{4\sqrt{n}g}{\omega}\right).
\end{equation}
completely in agreement with experimental results by Nakamura, Tsai and Pashkin\cite{naka1}. As also experimentally
observed, crossing at $m=n$ can happen belonging to interband transitions with $N=0$. Further details for
a single qubit case are given in Ref.\cite{fra1}.
At present, this is the only case where a comparison with experiment can be realized. 

\subsection{Two qubits}

For the case of two qubits one has
\begin{eqnarray}
\label{eq:h01}
    \tilde H_0 &=& \sum_n\frac{1}{2i}\sqrt{2}\Delta M_n\left(
    e^{-i\frac{4g^2}{\omega}t}
    \left|[n];1,1\right\rangle\left\langle [n];1,0\right|\right. \\ \nonumber
    &+&e^{i\frac{4g^2}{\omega}t}
    \left|[n];1,0\right\rangle\left\langle [n]; 1,-1\right| \\ \nonumber
    &-& e^{i\frac{4g^2}{\omega}t}
    \left|[n]; 1,0\right\rangle\left\langle [n]; 1,1\right| \\ \nonumber
    &-& \left.e^{-i\frac{4g^2}{\omega}t}
    \left|[n];1,-1\right\rangle\left\langle [n]; 1,0\right|
    \right).
\end{eqnarray}
Now, the Hamiltonian shows time-dependence and it is not straightforward to interpret the resonance conditions as
energy level crossings. On the same ground one has
\begin{eqnarray}
\label{eq:h11}
    \tilde H_1 &=& \sum_{\stackrel{m,n}{m\neq n}}e^{-i(n-m)\omega t}
    \frac{1}{2i}\sqrt{2}\Delta \left(
    e^{-i\frac{4g^2}{\omega}t}M_{mn}^{-}
    \left|[m];1,1\right\rangle\left\langle [n];1,0\right|\right. \\ \nonumber
    &+&e^{i\frac{4g^2}{\omega}t}M_{mn}^{-}
    \left|[m];1,0\right\rangle\left\langle [n]; 1,-1\right| \\ \nonumber
    &-& e^{i\frac{4g^2}{\omega}t}M_{mn}^{+}
    \left|[m]; 1,0\right\rangle\left\langle [n]; 1,1\right| \\ \nonumber
    &-& \left.e^{-i\frac{4g^2}{\omega}t}M_{mn}^{+}
    \left|[m];1,-1\right\rangle\left\langle [n]; 1,0\right|
    \right).
\end{eqnarray}
One can immediately diagonalize Hamiltonian (\ref{eq:h01}) obtaining eigenvalues $E_{n,s_x}=s_x\Delta M_n$
with $s_x=1,0,-1$. The corresponding eigenvectors are
\begin{eqnarray}
\label{eq:h1st}
     |n; 1, t\rangle &=& \frac{1}{2i}e^{-i\frac{4g^2}{\omega}t}|[n];1,1\rangle
                        +\frac{1}{\sqrt{2}}|[n];1,0\rangle
                        -\frac{1}{2i}e^{-i\frac{4g^2}{\omega}t}|[n];1,-1\rangle \\ \nonumber
     |n; 0, t\rangle &=& \frac{1}{\sqrt{2}}|[n];1,1\rangle+\frac{1}{\sqrt{2}}|[n];1,-1\rangle \\ \nonumber
     |n; -1, t\rangle &=&-\frac{1}{2i}e^{-i\frac{4g^2}{\omega}t}|[n];1,1\rangle
                        +\frac{1}{\sqrt{2}}|[n];1,0\rangle
                        +\frac{1}{2i}e^{-i\frac{4g^2}{\omega}t}|[n];1,-1\rangle
\end{eqnarray}
that do not give any geometrical phase as 
$\langle n; 1, t|i\partial_t|n; 1, t\rangle=
\langle n; 0, t|i\partial_t|n; 0, t\rangle=
\langle n; -1, t|i\partial_t|n; -1, t\rangle=0$. So, we can put again
\begin{equation}
\label{eq:ans1}
     |\psi(t)\rangle = \sum_{k,s_x}e^{-iE_{k,s_x}t}c_{k,s_x}(t)|k; s_x, t\rangle 
\end{equation}
and obtain the equations for the amplitudes
\begin{eqnarray}
\label{eq:amp1}
     i\dot c_{k,1}(t)&=& \sum_n 
     \frac{\Delta}{4}(M_{kn}^{-}+M_{kn}^{+})
     e^{-i[E_{n,1}-E_{k,1}+(n-k)\omega]t}c_{n,1}(t)+ \\ \nonumber
     & &\sqrt{2}\frac{\Delta}{4i}(M_{kn}^{-}-M_{kn}^{+})
     e^{i[E_{k,1}+\frac{4g^2}{\omega}-(n-k)\omega]t}c_{n,0}(t) \\ \nonumber
     i\dot c_{k,0}(t)&=& \sum_n 
     \sqrt{2}\frac{\Delta}{4i}(M_{kn}^{-}-M_{kn}^{+})
     e^{-i[E_{k,1}+\frac{4g^2}{\omega}+(n-k)\omega]t}c_{n,1}(t)+ \\ \nonumber
     & &e^{-i[E_{k,-1}+\frac{4g^2}{\omega}+(n-k)\omega]t}c_{n,-1}(t) \\ \nonumber
     i\dot c_{k,-1}(t)&=& \sum_n 
     -\frac{\Delta}{4}(M_{kn}^{-}+M_{kn}^{+})
     e^{-i[E_{n,-1}-E_{k,-1}+(n-k)\omega]t}c_{n,-1}(t)+ \\ \nonumber
     & &\sqrt{2}\frac{\Delta}{4i}(M_{kn}^{-}-M_{kn}^{+})
     e^{i[E_{k,-1}+\frac{4g^2}{\omega}-(n-k)\omega]t}c_{n,0}(t)
\end{eqnarray}
where we can recognize again the effect of the selection rules $\Delta s_x=0,\pm 1$ on the
permitted transitions, originating from the resonance conditions
\begin{eqnarray}
\label{eq:res1}
     E_{n,\pm 1}-E_{k,\pm 1}+(n-k)\omega &=& 0 \\ \nonumber
     E_{k,\pm 1}+\frac{4g^2}{\omega}-(n-k)\omega &=& 0
\end{eqnarray}
for intraband and interband transitions respectively,
giving rise to Rabi oscillations in the two-qubit system. Again, we can interpret this resonance
conditions as originating from the crossings of the energy levels for the Dicke model
\begin{equation}
     E_{n,s_x} = n\omega - \frac{4s_x^2g^2}{\omega} + s_x\Delta M_n
\end{equation}
that is degenerate with respect to $s_x=\pm 1$
but the degeneracy is removed by the last term. 
The Rabi frequencies 
can be computed from eqs.(\ref{eq:amp1}) with the rotating wave approximation imposing the resonance
conditions (\ref{eq:res1}) and
are given by, for intraband transitions
\begin{equation}
     {\cal R}_2 = {\cal R}_1
\end{equation}
and for interband transitions
\begin{equation}
     {\cal R'}_2 = \sqrt{2}{\cal R'}_1.
\end{equation}
These frequencies, in the limit of a large number of photons, displays dependence on the integer order
Bessel functions, being proportional to ${\cal R}_1$ and ${\cal R'}_1$
like in the single qubit case.
%

A cooperative effect arises from the coherent behavior of both the junctions entering entangled states and
producing Rabi oscillations.
%

So,
the main conclusion of this section is that the two qubits act collectively producing
Rabi oscillations, exactly as in the case of a single qubit. Such
oscillations appear between entangled states to be considered macroscopic.

\subsection{Three qubits}

In order to gain further insight into the physics of Josephson junctions in this case,
we analyze a system with three qubits. This case is rather different from the preceding
ones 
as $H_0$ depends on time and we have to apply a theorem for strong coupling that imposes
formally the adiabatic theorem at the leading order\cite{fra5}. So
%
there is a geometrical contribution to the phases. 
This makes no straightforward to
interpret the resonance conditions as energy level crossings. We have
\begin{eqnarray}
\label{eq:h032}
    \tilde H_0 &=& \sum_n\frac{1}{2i}\Delta M_n\left(
    e^{-i\frac{8g^2}{\omega}t}\sqrt{3}
    \left|[n];\frac{3}{2},\frac{3}{2}\right\rangle\left\langle [n];\frac{3}{2},\frac{1}{2}\right|\right. \\ \nonumber
    &+&2
    \left|[n];\frac{3}{2},\frac{1}{2}\right\rangle\left\langle [n];\frac{3}{2},-\frac{1}{2}\right| \\ \nonumber
    &-&e^{i\frac{8g^2}{\omega}t}\sqrt{3}
    \left|[n];\frac{3}{2},\frac{1}{2}\right\rangle\left\langle [n];\frac{3}{2},\frac{3}{2}\right| \\ \nonumber
    &+& e^{i\frac{8g^2}{\omega}t}\sqrt{3}
    \left|[n];\frac{3}{2},-\frac{1}{2}\right\rangle\left\langle [n];\frac{3}{2},-\frac{3}{2}\right| \\ \nonumber
    &-&2
    \left|[n];\frac{3}{2},-\frac{1}{2}\right\rangle\left\langle [n];\frac{3}{2},\frac{1}{2}\right| \\ \nonumber
    &-& \left. e^{-i\frac{8g^2}{\omega}t}\sqrt{3}
    \left|[n];\frac{3}{2},-\frac{3}{2}\right\rangle\left\langle [n];\frac{3}{2},-\frac{1}{2}\right|
    \right)
\end{eqnarray}
and
\begin{eqnarray}
\label{eq:h132}
    \tilde H_1 &=& \sum_{\stackrel{m,n}{m\neq n}}e^{-i(n-m)\omega t}\frac{1}{2i}\Delta \left(
    e^{-i\frac{8g^2}{\omega}t}\sqrt{3}M_{mn}^{-}
    \left|[m];\frac{3}{2},\frac{3}{2}\right\rangle\left\langle [n];\frac{3}{2},\frac{1}{2}\right|\right. \\ \nonumber
    &+&2M_{mn}^{-}
    \left|[m];\frac{3}{2},\frac{1}{2}\right\rangle\left\langle [n];\frac{3}{2},-\frac{1}{2}\right| \\ \nonumber
    &-&e^{i\frac{8g^2}{\omega}t}\sqrt{3}M_{mn}^{+}
    \left|[m];\frac{3}{2},\frac{1}{2}\right\rangle\left\langle [n];\frac{3}{2},\frac{3}{2}\right| \\ \nonumber
    &+& e^{i\frac{8g^2}{\omega}t}\sqrt{3}M_{mn}^{-}
    \left|[m];\frac{3}{2},-\frac{1}{2}\right\rangle\left\langle [n];\frac{3}{2},-\frac{3}{2}\right| \\ \nonumber
    &-&2M_{mn}^{+}
    \left|[m];\frac{3}{2},-\frac{1}{2}\right\rangle\left\langle [n];\frac{3}{2},\frac{1}{2}\right| \\ \nonumber
    &-& \left. e^{-i\frac{8g^2}{\omega}t}\sqrt{3}M_{mn}^{+}
    \left|[m];\frac{3}{2},-\frac{3}{2}\right\rangle\left\langle [n];\frac{3}{2},-\frac{1}{2}\right|
    \right).
\end{eqnarray}
It is straightforward to diagonalize Hamiltonian (\ref{eq:h032}) with the eigenstates
\begin{eqnarray}
    |n; s_x, t\rangle &=& \beta(s_x)\left[
    -ie^{-i\frac{8g^2}{\omega}t}\frac{\sqrt{3}}{2s_x}\left|[n];\frac{3}{2},\frac{3}{2}\right\rangle 
    +\left|[n];\frac{3}{2},\frac{1}{2}\right\rangle\right. \\ \nonumber
    &+&\left.i\left(s_x-\frac{3}{4s_x}\right)\left|[n];\frac{3}{2},-\frac{1}{2}\right\rangle
    -e^{-i\frac{8g^2}{\omega}t}\frac{\sqrt{3}}{2}\left(1-\frac{3}{4s_x^2}\right)
    \left|[n];\frac{3}{2},\frac{3}{2}\right\rangle
    \right]
\end{eqnarray}
being
\begin{equation}
    \beta(s_x)=\frac{1}{\sqrt{s_x^2+\frac{3}{16s_x^2}+\frac{27}{64s_x^4}+\frac{1}{4}}}
\end{equation}
and $s_x=\frac{3}{2},\frac{1}{2},-\frac{1}{2},-\frac{3}{2}$. As in the other cases, we get entangled
macroscopic states. In this case we have geometrical phases as given by
\begin{equation}
    \dot\gamma(s_x)=\frac{2g^2}{\omega}\frac{3-\frac{3}{2s_x^2}+\frac{27}{16s_x^4}}
    {s_x^2+\frac{3}{16s_x^2}+\frac{27}{64s_x^4}+\frac{1}{4}}
\end{equation}
that reduces to $\frac{2g^2}{\omega}$ for $s_x=\frac{3}{2},-\frac{3}{2}$ and $\frac{6g^2}{\omega}$ for $s_x=\frac{1}{2},-\frac{1}{2}$.

For the given $\tilde H_0$ and $\tilde H_1$ we seek a solution like
\begin{equation}
\label{eq:ans32}
     |\psi(t)\rangle = \sum_{k,s_x}e^{-iE_{k,s_x}t}e^{-i\dot\gamma_{s_x}t}c_{k,s_x}(t)|k; s_x, t\rangle 
\end{equation}
yielding the equations for the amplitudes
\begin{equation}
\label{eq:amp2}
    i\dot c_{m,\tilde s_x}(t)=\sum_{n,s_x}e^{-i[E_{n,s_x}-E_{m,\tilde s_x}+(n-m)\omega+\dot\gamma_{\tilde s_x}-\dot\gamma_{s_x}]t}\Delta\beta(s_x)\beta(\tilde s_x)\left[\alpha(s_x,\tilde s_x)M_{mn}^{+}+\alpha(\tilde s_x, s_x)M_{mn}^{-}\right]c_{n,s_x}(t)
\end{equation}
having set
\begin{equation}
   \alpha(s_x,\tilde s_x)=\frac{3}{4s_x}+\tilde s_x - \frac{3}{4\tilde s_x}+\frac{3}{4}\left(1-\frac{3}{4\tilde s_x^2}\right)\left(s_x-\frac{3}{4s_x}\right).
\end{equation}
It is not difficult to verify, also in this case, that the selection rules $\Delta s_x=0,\pm 1$ hold corresponding
to intraband $(s_x=\tilde s_x)$ and interband $(s_x\neq\tilde s_x)$ transitions respectively with the resonance conditions
\begin{equation}
\label{eq:res2}
E_{n,s_x}-E_{m,\tilde s_x}+(n-m)\omega+\dot\gamma_{\tilde s_x}-\dot\gamma_{s_x}=0. 
\end{equation}
The Rabi frequencies 
can be computed from eqs.(\ref{eq:amp2}) with the rotating wave approximation imposing the resonance
conditions (\ref{eq:res2}) and
can be given explicitly as follows
\begin{eqnarray}
     {\cal R}_{3/2,3/2}&=&{\cal R}_{-3/2,-3/2}=\frac{3}{2}{\cal R}_1 \\ \nonumber
     {\cal R}_{1/2,1/2}&=&{\cal R}_{-1/2,-1/2}=\frac{1}{2}{\cal R}_1
\end{eqnarray}
for intraband transitions and
\begin{eqnarray}
     {\cal R}_{3/2,1/2}&=&{\cal R}_{1/2,3/2}=\frac{\sqrt{3}}{2}{\cal R'}_1 \\ \nonumber
     {\cal R}_{-3/2,-1/2}&=&{\cal R}_{-1/2,-3/2}=\frac{\sqrt{3}}{2}{\cal R'}_1 \\ \nonumber
     {\cal R}_{1/2,-1/2}&=&{\cal R}_{-1/2,1/2}={\cal R'}_1
\end{eqnarray}
for interband transitions. Also for three qubit the Rabi frequencies are proportional to ${\cal R}_1$ and ${\cal R'}_1$, so in the limit of a large number of photons one has that these frequencies are given by integer order Bessel
functions. This rule applies to all cases.
 
As also seen for one and two qubits we can see again a collective effect. The behavior of the three qubits is
coherent producing entangled states with Rabi oscillations. Apart from numerical factors, the Rabi frequencies
are the same as in the single qubit case. It would be interesting to extend the analysis to higher spins to see if
this is a rule.


\section{\label{sec4} Thermodynamic limit: Quantum Amplifier}

To analyze this case, we limit the study to the leading order as higher orders become increasingly less important
as the number of Josephson junctions increases \cite{fra4}. This permits us to write down immediately the unitary
evolution being given by 
$H_0$ in
eq.(\ref{eq:uf}).
It is important to emphasize that, in order to unveil this effect, we should avoid random phases putting all the
junctions in the same state. This makes the case somewhat different from above but shows again a collective effect
possible to be observed.

Now, let us assume that the field in the cavity is in the ground state and {\em all the Josephson junctions are in their
lower state}
externally imposed.
Unitary evolution gives
\begin{equation}
    |\phi(t)\rangle = \sum_ n e^{-i\left(n\omega - \frac{g^2N^2}{\omega}\right)t}
    \left|[n]; \frac{N}{2}, -\frac{N}{2}\right\rangle e^{-\frac{N^2g^2}{\omega^2}}\left(\frac{Ng}{\omega}\right)^n\frac{1}{\sqrt{n!}}.
\end{equation}
So, leaving aside the contributions of the Josephson junctions, one is left with
\begin{equation}
    |N\chi(t)\rangle = e^{-\frac{Ng}{\omega}(a-a^\dagger)}\sum_ n e^{-i\left(n\omega - \frac{g^2N^2}{\omega}\right)t}
     e^{-\frac{N^2g^2}{\omega^2}}\left(\frac{Ng}{\omega}\right)^n\frac{1}{\sqrt{n!}}|n\rangle.
\end{equation}
that is, a coherent state with a parameter proportional to $N$. In the large $N$ limit this represents a classical state of the radiation field\cite{mand} and we see that we have amplified quantum fluctuations of the ground state of the radiation field to a classical level. The interesting aspect of this already known result is that it can be practically realized through Josephson junctions that lend themselves to realize this kind of classical state out of a fully quantum initial state. One realizes a QAMP (Quantum Amplifier)\cite{fra2,fra4}.

The device can be realized without too much care about taking just the limit $N\rightarrow\infty$ or $N\rightarrow\infty$, $V\rightarrow\infty$, being $V$ the volume of the cavity, and $N/V=constant$ as we are in the strong coupling limit.

This represents a very peculiar collective effect of a large number of Josephson junctions in a cavity in the strong coupling regime,
the same regime
devised in the experiments of Nakamura's group\cite{jj1,jj2}
for a single qubit.

\section{\label{sec5} Discussion and Conclusions}

The above discussed collective effects can have their limit in the appearance of decoherence. The experiments of
Nakamura's group display the decay of Rabi oscillation but the decay time can be enough long to permit quantum computation. Neither the nature of this decoherence effects nor the way it scales with the number of Josephson junctions are 
%
known as far as we know. It should be expected that the same should happen for the QAMP producing a classical radiation field. The origin of decoherence on Josephson junctions is to be understood but it could not be excluded that revival and collapse effects of Rabi oscillations may be at work as happens in the weak coupling regime\cite{stro1,stro2}.

We have shown how, in the strong coupling limit as devised by Nakamura's group, several collective effects as Rabi oscillations and quantum amplification could appear in Josephson junctions coupled by a radiation field. Rabi oscillations, in the limit of a large number of photons, have frequencies proportional to integer order Bessel functions and a selection rule applies limiting
%
transitions. This effects could be useful for applications in quantum computation or to use a large number of Josephson junctions to produce a laser field.

\end{document}